\documentclass[apj,twocolumn]{emulateapj}
\usepackage{amsmath}
\usepackage{txfonts}
\eqsecnum

\renewcommand\email\texttt

\begin{document} 

\slugcomment{\sc submitted to \it the Astrophysical Journal}
\shorttitle{\sc The Discovery of Tidal Tails Around NGC 5466}
\shortauthors{\sc V.~Belokurov et al.}

\title{The Discovery of Tidal Tails Around the Globular Cluster NGC
5466} \author{V. Belokurov, N.W. Evans, M.J. Irwin, P.C. Hewett \&
M.I. Wilkinson} \affil{Institute of Astronomy, University of
Cambridge, Madingley Road, Cambridge CB3 0HA, UK}

\begin{abstract}
We report the discovery of tidal tails around the high-latitude
Galactic globular cluster NGC 5466 in Sloan Digital Sky Survey (SDSS)
data. Neural networks are used to reconstruct the probability
distribution of cluster stars in $u,g,r,i,z$ space.  The tails are
clearly visible, once extra-galactic contaminants and field stars have
been eliminated.  They extend $\sim 4^\circ$ on the sky, corresponding
to $\sim 1$ kpc in projected length. The orientation of the tails is
in good agreement with the cluster's Galactic orbit, as judged from
the proper motion data.
\end{abstract}


\section{Introduction}

The description of the Milky Way Galaxy in terms of two simple
components -- Population I and Population II stars -- is now primarily
of historical interest. Rather, the Galaxy has extensive substructure
in both configuration and velocity space caused by the merging and
accretion of satellite galaxies~\citep{Ib97,Ma03,Ya03} and the
disruption of star clusters~\citep{Le00,Od01}. 


Tidal tails around globular clusters provide a good example of such
substructure. Large, multi-color fields of some of the Galactic
globular clusters are readily available (for example, through the
Sloan Digital Sky Survey, SDSS). The difficulty is to distinguish
between tidal tail stars and field stars with the help of
(generalizations of the) color-magnitude diagrams and to choose an
analysis technique that maximises the signal-to-noise ratio of the
surface density of the tidal tail.

There are 10 Galactic globular clusters imaged by SDSS, namely NGC
2419, 5272, 5466, 6205, 7078, 7089 and Palomar 3, 4, 5, 14.  Of these,
the lowest mass ones are Pal 3, 4, 5, 14 and NGC 5466~\citep{Ha96}.
Pal 3, 4 and 14 lie at Galactocentric distances of $\sim 70$ to 110
kpc and so are likely to be less battered by the Galactic tides. Pal 5
has already been the subject of extensive investigations~\citep{Od03},
leaving NGC 5466 as the next obvious candidate. There has already been
a claim of possible evidence for a tidal perturbation in the outer
parts of NGC 5466 in the APM scans of POSS I plates~\citep{Od04}. In
this {\it Letter}, we analyze the SDSS data on NGC 5466 in detail and
demonstrate the existence of extensive tidal tails.

\begin{figure*}[t]
\includegraphics[height=7.5cm]{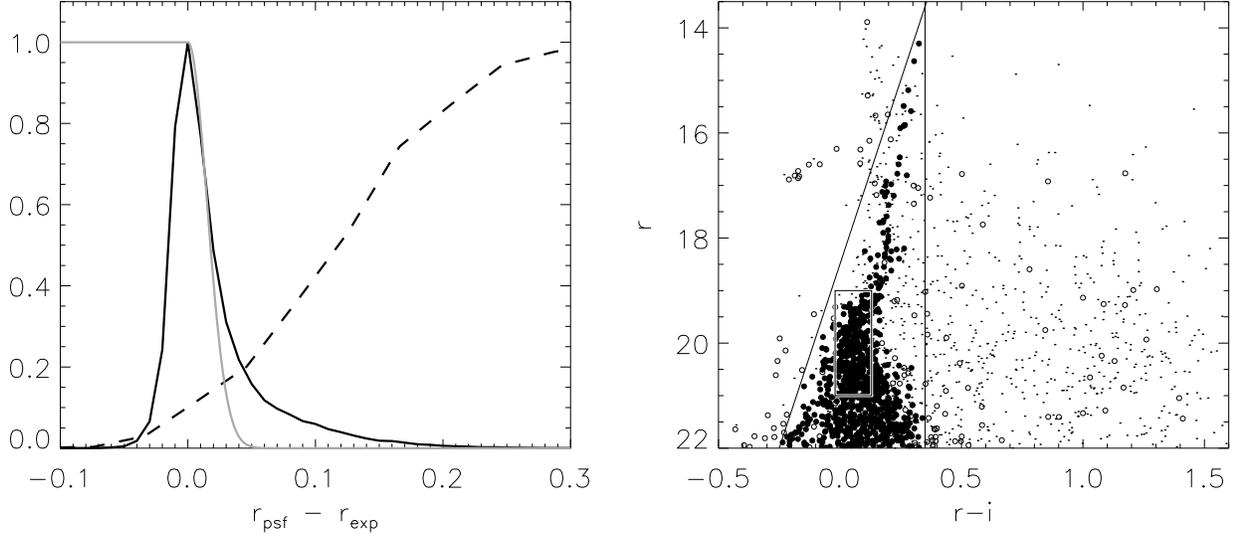}
\caption{Left: Histogram of the concentration index $r_{\rm psf} -
r_{\rm exp}$ for objects classified by SDSS as stars (full black line)
and galaxies (broken line) in the field of view. Each distribution has
been normalized to have peak value unity (see Fig. 2 of Scranton et
al. (2002) for the unnormalized distribution of all objects). The full
grey line shows the taper applied to scale down the contribution from
the objects that are likely to be background galaxies (see Fig 3 of
Lupton et al. (2001), for the misclassification rate as a function of
magnitude). Right: The color-magnitude diagram showing cluster members
(filled circles) and field stars (dots) in the training set. The
unfilled circles are rejected by the color cuts shown as solid lines
or the subsidiary cut in $g-r$, and are not used in the training
set. The rectangular box is used to set the overall normalisation, as
described in the main text.}
\label{fig:cmds}
\end{figure*}
\begin{figure*}
\includegraphics[width=\hsize]{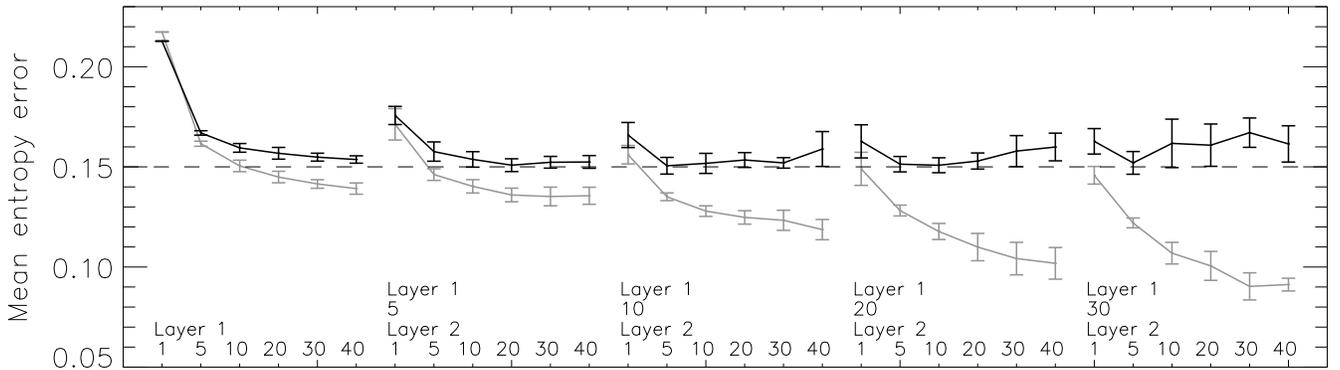}
\caption{Plots of the mean entropy error in the training set (grey)
and test set (black) for a variety of network architectures. The
leftmost example has just one hidden layer, the remaining two examples
have hidden layers. The numbers of neurons are given in the figure.}
\label{fig:nns}
\end{figure*}

\section{Method}

The detection of tidal tails is really a two-class problem. We want to
know whether a given star belongs to the cluster or the field.  An
optimal contrast or matched filter approach (e.g., Rockosi et
al. 2002, Odenkirchen et al. 2003) works in the following way. It
constructs the two class conditional probabilities $P(x|C_1)$ (for the
cluster) and $P(x|C_2)$ (for the field stars). Here, $x$ denotes the
data, which can either be raw magnitudes or color indices. The class
conditional probabilities can be easily constructed from
two-dimensional data histograms, given regions almost exclusively
composed of either cluster stars or field stars.  In a least squares
solution such as that of Odenkirchen et al. (2003), the class
probability $P(C_1)$ is proportional to the ratio of
$P(x|C_1)/P(x|C_2)$ summed over all stars in any given bin in right
ascension and declination. This can be translated into the surface
density of cluster stars.

\citet{Od03} used this technique successfully on two-dimensional data
comprising the $i$ band magnitude and the color index $c_1 = 0.91(g-r)
+ 0.42(r-i)$.  For higher dimensional data, simple binning will not
work. For example, if we have $\sim 1000$ cluster members, most bins
will be empty. This is partly why \citet{Od03} compressed the data by
constructing a color index.

An alternative technique is as follows. {\it Multi-layer
perceptrons}~\citep{Bi95} can approximate the posterior class
probability $P(C_1|x)$ in a high dimensional data-space with a limited
number of datapoints. Of course, a training set of cluster stars and
background stars is still needed. It can be provided by taking stars
close to the center of the cluster, which have a target value of
unity, and far away from the cluster, with target value of zero.  Once
the neural network is trained, the fraction of cluster stars in a
given right ascension and declination bin is just the mean output of
the network
\begin{equation}
P(C_1) \approx {1 \over N} \sum_{i=1}^N P(C_1|x)
\label{eq:firsteq}
\end{equation}
where $N$ is the number of stars in the bin. 

\section{Application to NGC 5466}

SDSS imaging data are obtained almost simultaneously in five
photometric bands namely, $u, g, r, i$, and $z$~\citep{Fu96,Yo00}.
The sky coverage~\footnote{see
http://www.sdss.org/dr4/dr4photogal\_big.gif} of Data Release 4
(Adelman-McCarthy et al. 2005) is 6670 square degrees, mostly at high
Galactic latitude.  This is a homogeneous dataset with excellent
photometric accuracy, but in dense star fields, like the cores of
globular clusters, the pipeline suffers from crowding problems (see
e.g., Lupton et al. 2001).

NGC 5466 is a metal-poor Galactic globular cluster that lies $\sim 16$
kpc from the Sun. It has (J2000) equatorial coordinates ($\alpha =
211.3^\circ, \delta = 28.5^\circ$), which correspond to galactic
coordinates ($\ell =42.2^\circ, b=73.6^\circ$).  We use a magnitude
limited ($14.0 < r_{\rm psf} < 22.0$) object catalogue, where $r_{\rm
psf}$ is the point spread function magnitude corrected for extinction
using the Schlegel et al. (1998) maps. This leaves us 104016 objects
classified as stars in the field $6^\circ \times 6^\circ$ around the
center of the cluster.

Visual examination of APM-scanned distributions of POSS II detected
images in this region already reveals discernible galaxies in the
vicinity of the cluster. SDSS provides a crude separation into stars
and galaxies, but there is a contamination of the stellar sample by
galaxies, particularly at faint magnitudes.  The left panel of
Figure~\ref{fig:cmds} shows the difference between the magnitudes
obtained by PSF photometry and by fitting an exponential profile,
namely $r_{\rm psf} - r_{\rm exp}$.  This is the concentration
parameter introduced by Scranton et al. (2002). Stars are tightly
concentrated around zero, whereas galaxies show a significant positive
excess. We minimize contamination by external galaxies by using the
taper function plotted as a solid grey line in Figure~\ref{fig:cmds}.
The taper is unity for $r_{\rm psf} - r_{\rm exp} < 0$ and is one wing
of a Gaussian with dispersion $0.015$ for $r_{\rm psf} - r_{\rm exp}>
0$. This taper function is used to multiply the output probabilities
of the neural networks, and therefore gives less weight to the
contribution of likely contaminants. Effectively, this means that we
only use stars with $r_{\rm psf} - r_{\rm exp} < 0.05$, which number
87760 in total.

We select cluster stars to lie within a distance $r_1 = 0.14^\circ $
of the cluster center. We have chosen $r_1$ to be larger than the
half-light radius ($0.038^\circ$) because the central parts of the
cluster are missing.  We identify the upper main sequence, sub-giant
and giant branch stars from the color cuts, $r-i < 0.35$ and $r > -14
(r-i) + 18.5$. There is also a subsidiary color cut~\footnote{This cut
encloses the cluster sequence in $g-r, r$ and removes a small number
of obvious outliers, but otherwise has no effect on our results} in
$g-r$ to refine the selection, which leaves 896 stars. These are shown
in the right panel of Figure~\ref{fig:cmds} as filled circles. The
open circles are the remaining stars within $r_1$, predominantly
foreground stars and some horizontal branch stars belonging to the
cluster.  The filled circles are the stars used in the training set
with target value unity. The dots are stars lying beyond $r_2 =
1^\circ$, larger than any estimate of the tidal radius of the cluster.
We choose these stars to have target value zero. They are mainly
field stars belonging to the disk and the halo.

Ideally, we would like the region in the data space around the
decision boundary to be well-sampled in the training set.  The number
of selected stars is set by requiring the field stars and the cluster
stars around the cluster main sequence turn-off to be roughly equal.
In practice, this is implemented by requiring the training set to have
roughly equal numbers of the populations in the color-magnitude box
$-0.02 <r-i < 0.13$ and $19< r < 21$. This gives a total of $\sim
7000$ stars in the training set.

In a multi-layer perceptron, the neurons or processing units are
arranged in layers. The input data are fed to the bottommost
layer. The output value emerges from the topmost layer, the
intervening layers are hidden.  The value of the neuron in any layer
is calculated by the sum over the weights multiplied by the
activations of the neurons connected to it. The activation is computed
from the value of the neuron via an activation function. We use the
logistic function $f(v) = 1/(1 + \exp(-v))$, where $v$ is the value of
the neuron (Belokurov et al. 2003). This allows us to interpret the
output of the network as a posterior probability (Bishop 1995, section
6.7).  Given the input data and a set of weights, we can construct an
error function, which quantifies the performance of the network. We
use a particular form appropriate for two-class problems, namely the
cross-entropy error (Bishop 1995, section 6.10).  During training, we
obtain the weights that minimise the error function over the training
set using a steepest descent scheme. In practice, this is done by
performing a sequence of iterative up-dates using a variant of
back-propagation, which helps to prevent entrapment in local minima
(see e.g., Bishop 1995, section 7.5). A further refinement is to add
an additional term to the error function, which is the sum of the
squares of the weights multiplied by a weight decay
parameter. Adjusting this parameter enables us to control the
magnitude of weights and hence to minimise any over-fitting. This can
be done automatically during training.

The neural networks are constructed with the Stuttgart Neural Network
Simulator (SNNS). There are five neurons in the input layer,
corresponding to the normalized fluxes in $u,g,r,i$ and $z$. By
normalized, we mean the following. The mean and dispersion of the
fluxes in the entire field are computed. The network input is then the
flux with the mean subtracted and scaled by the standard deviation
(e.g., Belokurov et al. 2003). There is one output layer, which
approximates the probability of a globular cluster star given the data
$P(C_1|x)$.  The training algorithm is resilient back-propagation with
adaptive weight decay (Bishop 1995, section 7.5.3). The networks are
trained for a total of 4000 iterations, with weight decay parameter
adjusted every 200.

The number of hidden layers and neurons can be freely chosen and is
set by experimentation on the data.  Figure~\ref{fig:nns} shows the
mean cross-entropy error -- defined as the total error normalised by
the number of degrees of freedom (number of patterns minus number of
network parameters) -- for the training and test sets for different
network architectures. In this experiment, the original training set
was split randomly into two of equal sizes. One of these only was used
for training, the other for testing. Of course, there are no patterns
in common. As discussed by Belokurov et al. (2004), the error in the
training set diminishes with increasing network complexity, but the
error in the test set flattens off and eventually starts to rise.  An
optimum architecture is therefore given by the onset of flattening, as
more neurons in the hidden layers will make little further
improvement. Examination of the figure suggests that networks with
($5:5:20:1$) or ($5:10:5:1$) or ($5:20:5:1$) are optimal. This
notation refers to the number of neurons in the input, first hidden,
second hidden and output layers.  For our final results, we use
committees of 10 neural networks with architecture ($5:10:5:1$),
averaging over the network outputs.

\begin{figure}[t]
\begin{center}
\includegraphics[height=5.5cm]{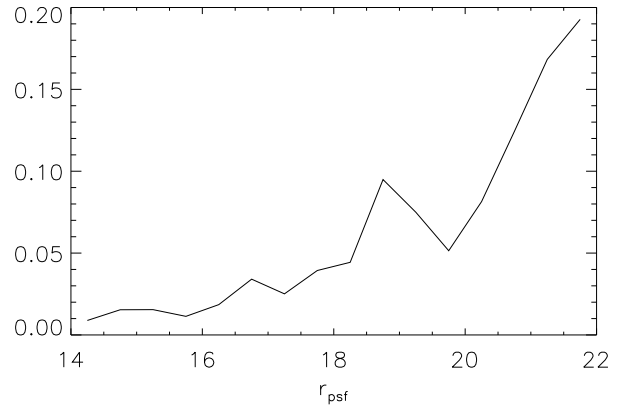}
\caption{Plot of the ratio of mean outputs for field and for
globular cluster stars as a function of magnitude bin $r_{\rm psf}$.}
\end{center}
\label{fig:fps}
\end{figure}
\begin{figure*}[t]
\includegraphics[height=8.5cm]{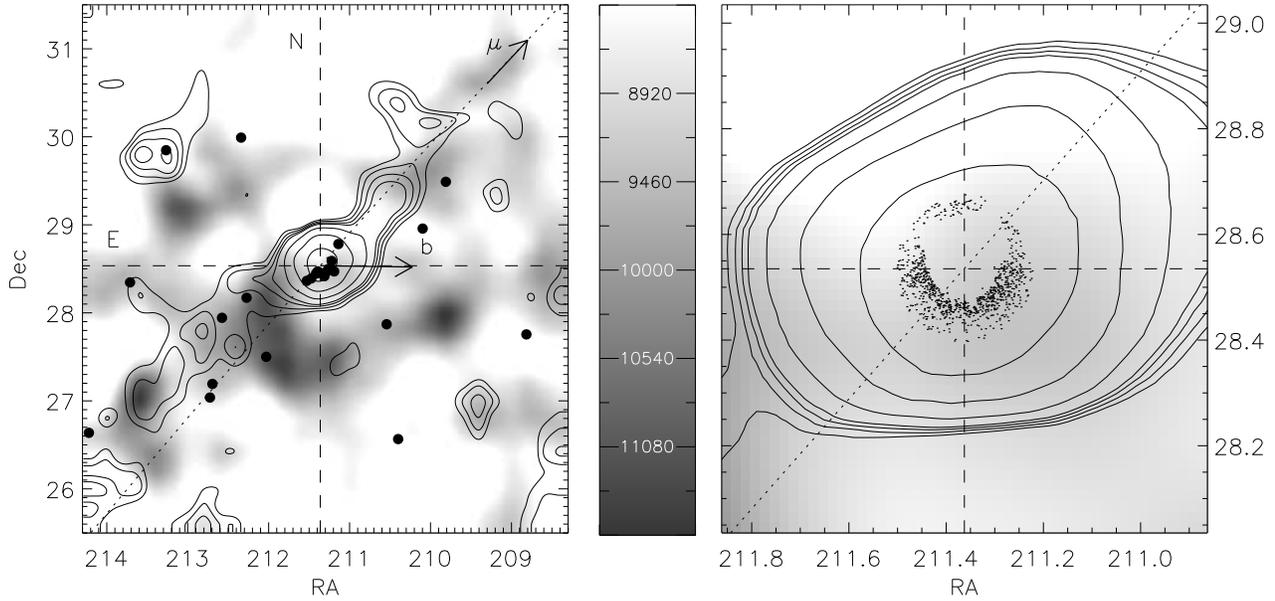}
\caption{Left: Contours of mean network output (proportional to the
star density) for the globular cluster NGC 5466.  The contours mark
the levels with network output at $1.5, 2, 2.5, 3, 5, 10$ and $20
\sigma$ above the mean (see text for details). The trajectory is shown
as a dotted line.  An arrow marked with $\mu$ shows the direction of
motion, while one marked with $b$ shows the direction of increasing
Galactic latitude. The grey-scale shows the background galaxy density
(in number per square degree). Filled circles are blue horizontal
branch stars. Right: Detail of the innermost $1^\circ \times 1^\circ$
of the cluster. The dots are the stars used in the training set. [The
left (right) panel uses Gaussian smoothing with FWHM $= 0.3^\circ$
($=0.1^\circ$) and a median filter with window size $0.4^\circ$
($0.2^\circ$).]}
\label{fig:tail}
\end{figure*}

Figure~\ref{fig:fps} gives an idea of the effectiveness of the
classification. It shows the ratio of mean outputs for field and for
globular cluster stars as a function of magnitude bin.  For a perfect
classification, this ratio is close to zero. We see from
Figure~\ref{fig:fps} that the slope of the curve changes abruptly at
$r \approx 20$. This can be attributed to the broadening of the top of
the main sequence in color space. This increases the number of
misclassifications -- that is, the number of cluster stars assigned a
lower output and field stars assigned a higher output. This suggests
applying a magnitude taper function to the network outputs according
to the value of $r_{\rm psf}$. Specifically, the taper is unity for
stars brighter than $r_{\rm psf} = 20$, and is one wing of a Gaussian
with dispersion $0.4$ for stars fainter than $r_{\rm psf} =20$.

An alternative to using taper functions in concentration index and
magnitude is a hard cut, and tests show that this yields very similar
final results.

\section{Results}

To produce contour maps of the structure around NGC 5466, we first
divide the $6^\circ \times 6^\circ$ field of view into 900 bins with
size $0.2^\circ$ on a side. We then use Gaussian smoothing and median
filtering (the IDL routine {\tt FILTER\_IMAGE} as detailed in the
figure caption), followed by cubic splines for interpolation. This
yields the maps shown in Figure~\ref{fig:tail}, which show curves of
constant mean network output [eqn~(\ref{eq:firsteq})]. This is
directly proportional to the stellar number density. The contour
levels are chosen in the following way. The distribution of mean
network output $P(C_1)$ for all bins in the field of view is a
Gaussian with the centre close to zero and an extended right-hand
tail, corresponding to cluster stars.  The contour levels shown in
Figure~\ref{fig:tail} correspond to $P(C_1)$ values at $1.5, 2, 2.5,
3, 5, 10$ and $20 \sigma$ above the mean.

The inner part of the cluster is probably spherical, even though the
right panel of Figure~\ref{fig:tail} shows some distortion of the
inner contours.  This is understandable, as the SDSS pipeline suffers
from crowding problems in the dense center. This is illustrated by the
dots in the right panel, which show the cluster stars actually used in
the training set.

The outer parts of the cluster are deformed. The tidal tails of NGC 5466
are clearly visible, stretching out in a broadly symmetrical fashion
on both sides of the cluster. Also shown is the background galaxy
density, using the distribution of objects identified as galaxies by
the SDSS pipeline. There is a cluster of galaxies lying behind the
southern tail and this may partly account for the broken cluster
density contours, as well as some of the asymmetries in the
substructure between the tails.

There are two straightforward tests that support the tidal tail
hypothesis. First, the proper motion of NGC 5466 is known from {\it
Hipparcos} (Odenkirchen et al. 1997). The projection of the orbit is
plotted on Figure~\ref{fig:tail}. It is in good agreement with the
location of the tails, which should of course be distended along the
orbit. The position angle (with respect to Galactic North through
East) of the tails varies from $\sim 30^\circ$ at 20' to $\sim 45^\circ$ in
the tails. This is in good agreement with the tidally-induced
perturbations in the starcounts already found by Odenkirchen \& Grebel
(2004) using APM scanned POSS I plates of the inner 20' of NGC 5466.
Second, the positions of blue horizontal branch stars are also shown
in the Figure~\ref{fig:tail} as filled circles. Let us recollect the
horizontal branch was excised from the training set (see the right
panel of Figure~\ref{fig:cmds}) and so the positions of these stars
have not therefore contributed to the analysis so far. Nonetheless,
these stars also seem to form an extended distribution stretched out
along the tails, together with some field star interlopers. The total
number of blue horizontal branch stars is small (26), so this test is
suggestive rather than conclusive.

\section{Discussion and Conclusions}

We have identified tidal tails around the high latitude Galactic
globular cluster NGC 5466. There have been tentative indications of
tidal tails before from the starcounts in the inner $20'$ of the
cluster (Odenkirchen \& Grebel 2004). Here, we have provided an
unambiguous identification of extended structure around the
cluster. The tidal tails have been traced for over $4^\circ$ of arc,
using the high quality Sloan Digital Sky Survey (SDSS) photometric
data. Numerical simulations of the evolution of NGC 5466 are underway.

\acknowledgments

Funding for the creation and distribution of the SDSS Archive has been
provided by the Alfred P. Sloan Foundation, the Participating
Institutions, the National Aeronautics and Space Administration, the
National Science Foundation, the U.S. Department of Energy, the
Japanese Monbukagakusho, and the Max Planck Society. The SDSS Web site
is http://www.sdss.org/.

\end{document}